\documentclass{article}
\usepackage{amssymb}
\usepackage{amsmath}

\begin{document}
\centerline{\Large \bf The rational generalized integrating factors}
\medskip
\centerline{\Large \bf for first-order ODEs}

\vskip 2cm
\centerline{\sc Yu. N. Kosovtsov}
\medskip
\centerline{79060, 43-52, Nautchnaya Street, Lviv, Ukraine.}
\centerline{email: {\tt kosovtsov@escort.lviv.net}}
\vskip 1cm
\begin{abstract}
We describe a solving semi-decision method based on examination of the rational structures of the generalized integrating factors of first-order ODEs. We propose a conjecture that for some family of equations of the type $dy/dx=P(x,y)/Q(x,y)$, with $P$ and $Q$ polynomials only in $y$ (or in $x$), the general form of the structures of  generalized integrating factors are rational in $y$ (or in $x$). In such a way one can obtain a differential-algebraic polynomial system for undetermined parameters of the structures. The successful solution of this system (it is sufficient to find any particular solution) automatically leads to finding the general solutions of ODEs.
\end{abstract}

\section{Introduction}

Equations of the type

\begin{equation}
\frac{dy(x)}{dx}=f(x,y),
\label{ode}
\end{equation}
where $f(x,y)=P(x,y)/Q(x,y)$ with $P$ and $Q$ polynomials in $y$ arise in many different physical fields. There are some methods for finding the solutions for such type of equations. First of all it is the famous method of integrating factor. A remarkable method based on the knowledge of the general structure of the integrating factor was developed by Prelle and Singer \cite {Singer1} and Singer \cite {Singer2} for the case when $P$ and $Q$ are polynomials in \emph {both} arguments $x$ and $y$.

There are extensions of this method in some directions \cite {MacCallum}-\cite {Duarte2} including approaches to solve ODEs with some transcendental or algebraic terms (heuristically)) \cite {Shtokhamer}-\cite {Duarte3}.

Roughly speaking, the Prelle-Singer type methods consist of three parts. On first of them it is determined (or supposed) the general structure of the integrating factors for some family of ODEs or for some type (elementary, algebraic, Liouvillian and so on) of solutions. On the second part for given rational ODE one have to choose the more specific structure of integrating factor from its general form (by specifying degree bounds for polynomials in rational, exponential and so on parts of integrating factor). And then on the third part one have to try to find the undetermined parameters of conjectural structure.

The distinctive feature of the Prelle-Singer type of methods is that (except some relatively simple cases) we are not able theoretically define a degree bounds for polynomials of the structures of integrating factors, so we can not miss vexatious second step.

In present paper we propose an approach based on examination of structures of generalized integrating factors of equations with $P$ and $Q$ polynomials \emph {only} in $y$ (or \emph {only} in $x$).

The paper is organized as follows. In section 2 we introduce the generalized integrating factors. We note in Section 3 that rational hypothesis describes not only rational equations. Section 4 is devoted to consideration of general structures of the generalized integrating factors for rational ODEs. Here we introduce "Order guides" which is helpful tool for defining of procedure parameters. The important moment of the method - simplification of the system for unknown coefficients is discussed in Section 5. In Section 6 we give formulae to express the equation solutions through the generalized integrating factor.  One of the ways to generalize structures for rational ODEs is considered in Section 7.  In sections 8 and 9 we briefly touch on the cases when $P$ and $Q$ polynomials in $x$ only, and in both $x$ and $y$.

\section{Generalized integrating factors}

In \cite {Kosovtsov} we demonstrated on examples the solving abilities of the method based on structures of solutions of linear first-order PDE associated with (\ref{ode})

\begin{equation}
\frac{\partial \zeta (x,y)}{\partial x}+ f(x,y)\frac{\partial \zeta (x,y)}{\partial y}=0.
\label{pde}
\end{equation}
We examined there the structures of the type
\begin{equation}
\zeta (x,y)=\frac{p_0 (x,y)}{q_0 (x,y)}+\sum_i \,\alpha_i \ln \frac{p_i (x,y)}{q_i (x,y)} ,
\label{form}
\end{equation}
or equivalently
\begin{equation}
\tilde{\zeta} (x,y)=\exp(\frac{p_0 (x,y)}{q_0 (x,y)})\prod_i \, (\frac{p_i (x,y)}{q_i (x,y)})^{\alpha_i} ,
\label{eform}
\end{equation}
where $p_0$, $q_0$, $p_i$, $q_i$ are polynomials in $y$; $\alpha_i$ are constants.
The structures of type (\ref{form}), (\ref{eform}) are complicated enough. They may have many terms and we have to look over many preliminary propositions on the second step in choosing orders of polynomials $p_0$, $q_0$, $p_i$, $q_i$ and $\alpha_i$.

But from (\ref{form}) we can note that this structure is the result of integration of a rational function in $y$. It brings us an idea that we can find such characteristic, which lead to easy but more general structures and which has solvable connection at least with $\zeta (x,y)$.

It is easy to see that for classical integrating factor $\mu_y=\frac{\partial \zeta}{\partial y}$, which is determined by the following equation
\begin{equation}
\frac{\partial \mu_y}{\partial x}+ \frac{\partial f\mu_y}{\partial y}=0,
\label{muy}
\end{equation}
the possible structures have the same form as (\ref{eform}) although here they circumscribe a more rich ODEs family.

The same structures has the second integrating factor $\mu_x=\frac{\partial \zeta}{\partial x}$ with the following equation
\begin{equation}
\frac{\partial }{\partial x}\frac{\mu_y}{f}+ \frac{\partial \mu_y}{\partial y}=0.
\label{mux}
\end{equation}

Let us now introduce \emph{new} characteristic
\begin{equation}
\mu_{yy}=\frac{\partial}{\partial y}\ln \frac{\partial \zeta}{\partial y}.
\label{muyy}
\end{equation}
We will call such type of characteristics as generalized integrating factors.

The $\mu_{yy}$ is determined by the following \emph{non-homogeneous} equation
\begin{equation}
\frac{\partial \mu_{yy}}{\partial x}+ \frac{\partial f\mu_{yy}}{\partial y}+\frac{\partial^2 f}{\partial y^2}=0.
\label{muyyeq}
\end{equation}
Here we can see that \emph{for some rational $f(x,y)$ in $y$ there exist a particular solution for $\mu_{yy}$ in form of rational function in $y$}.

If we know \emph{any particular} solution for $\mu_{yy}$ we can find both integrating factor $\mu_y$ and PDE solution $\zeta$ and finally the general solution of equation (\ref{ode}) $y(x)$ (see Section 6).

It is obvious, that rational conjecture for $\mu_{yy}$ gives as particular cases all forms, considered above, for $\mu_y$ and $\zeta$.

The \emph{same} observation may be done for the following generalized integrating factor
\begin{equation}
\mu_{yx}=\frac{\partial}{\partial y}\ln \frac{\partial \zeta}{\partial x}.
\label{muyx}
\end{equation}

The $\mu_{yx}$ is determined by the following equation
\begin{equation}
\frac{\partial \mu_{yx}}{\partial x}+ \frac{\partial f\mu_{yx}}{\partial y}+\frac{\partial \ln f}{\partial x}\frac{\partial \ln f}{\partial y}-\frac {1}{f}\frac{\partial^2 f}{\partial x\partial y}=0.
\label{muyxeq}
\end{equation}

And also for
\begin{equation}
\mu_{xy}=\frac{\partial}{\partial x}\ln \frac{\partial \zeta}{\partial y}.
\label{muxy}
\end{equation}

The $\mu_{xy}$ is determined by the following equation
\begin{equation}
\frac{\partial} {\partial x}\frac{\mu_{xy}}{f}+ \frac{\partial \mu_{xy}}{\partial y}-\frac{\partial \ln f}{\partial x}\frac{\partial \ln f}{\partial y}+\frac{1}{f}\frac{\partial^2 f}{\partial x\partial y}=0.
\label{muxyeq}
\end{equation}

And at last for
\begin{equation}
\mu_{xx}=\frac{\partial}{\partial x}\ln \frac{\partial \zeta}{\partial x}.
\label{muxx}
\end{equation}

The $\mu_{xx}$ is determined by the following equation
\begin{equation}
\frac{\partial }{\partial x}\frac{\mu_{xx}}{f}+ \frac{\partial \mu_{xx}}{\partial y}+\frac{\partial^2 }{\partial x^2}\frac{1}{f}=0.
\label{muxxeq}
\end{equation}

Each of the generalized integrating factors can be expressed through another one and $f$ just as $\mu_x=-f\mu_y$. For example,
\begin{equation}
\mu_{yx}=\mu_{yy}+\frac{\partial \ln f}{\partial y},
\label{muyxyy}
\end{equation}

\begin{equation}
\mu_{xy}=-f\mu_{yx}
\label{muxyyx}
\end{equation}
and
\begin{equation}
\mu_{xx}=-f\mu_{yy}+\frac{\partial \ln f}{\partial x}-f \frac{\partial \ln f}{\partial y}.
\label{muxxyy}
\end{equation}
Then for $f$ rational if one of generalized integrating factor is  rational the rest can be choosen rational too.

We can conclude that:

{\bf Proposition:} \emph{For some family of first-order ODEs with rational $f(x,y)$ in $y$ with coefficients in the complex field $\mathbb{C}(x)$ there exists a particular solution of (\ref{muyyeq}) for $\mu_{yy}$ in form of rational function in $y$ with coefficients in $\mathbb{C}(x)$. The same is valid for $\mu_{yx}$, $\mu_{xy}$ and $\mu_{xx}$.}

\section{Some remarks on ODE families determined by rational generalized integrating factor}

If we solve, for example, equation (\ref{muyyeq}) with respect to $f$, we obtain that
\begin{equation}
f(x,y)=[F1(x)+\int(F2(x)-\int \frac{\partial \mu_{yy}}{\partial x}dy)\,\exp(\int \mu_{yy}dy) dy]\,\exp(-\int\mu_{yy}dy),
\label{fmuyy}
\end{equation}
where $F1(x)$ and $F2(x)$ are arbitrary functions.

It is easy to see here that for arbitrary rational $\mu_{yy}$ $f$ may not be an rational function. Even in simplest case when $\mu_{yy}=const$ $f$ is not rational.

When we consider structures for $\zeta$ in form (\ref{form}) we always receive $f$ rational but not all rational $f$ are satisfied by such a form. Analogous structures for $\mu_y$ or $\mu_x$ no longer give only rational $f$'s but they are not complete all rational $f$ too. We hope that rational generalized integrating factor can complete \emph{more} rational $f$.

From another side nothing prevent us to try to find solutions of equations (\ref{muyyeq}), (\ref{muyxeq}), (\ref{muxyeq}) and (\ref{muxxeq}) with non-rational $f$ starting from hypothesis that respective generalized integrating factor is rational.

It is interesting to note, for example, that the "degenerate" case when $\mu_{xy}=0$ ( or $\mu_{yx}=0$) corresponds to "separable" ODEs.

In the sequel however we reduce ourselves to consideration of rational $f$'s only.

\section{Order guides for rational generalized integrating factors}

Let now $f(x,y)=P(x,y)/Q(x,y)$ and $\mu_{yy}(x,y)=X(x,y)/Y(x,y)$, where $P$, $Q$, $X$ and $Y$ are polynomials in $y$. Here we have to consider $P$ and $Q$ as specified in advance, and $X$ and $Y$ as unknowns. That is we are going to seek $\mu_{yy}$ in the form of
\begin{equation}
\mu_{yy}(x,y)=\frac{X(x,y)}{Y(x,y)}=\frac{\sum_{i=0}^{N1} a1_i(x)y^i}{\sum_{i=0}^{N2} a2_i(x)y^i}
\label{HXYyy}
\end{equation}
with unknowns $a1_i(x)$ and $a2_i(x)$, $N1$ and $N2$. Then from equation (\ref{muyyeq}) we obtain that $X$ and $Y$ have to satisfy the following equation (partial derivatives are denoted by standard subscripts)
\begin{align}
Y^2(PQQ_{yy}-2PQ_y^2&-P_{yy}Q^2+2P_yQQ_y)+Q^3(XY_x-X_xY)+\notag \\&+PQ^2(XY_y-X_yY)+QXY(PQ_y-P_yQ)=0.
\label{XYyy}
\end{align}
It is a polynomial equation of the type
\begin{equation}
\sum_{i=0}^N A_i(x)y^i =0,
\label{Polyn}
\end{equation}
where $A_i(x)$ depend on polynomials coefficients and their derivatives. This equation disintegrates on a polynomial differential-algebraic system of $N+1$ equations for polynomial coefficients $\{A_i(x)=0\}$, which determines the "auxiliary" system for unknown coefficients of $X$ and $Y$. To solve this system we first of all have to capture orders of unknown polynomials $X$ and $Y$.

Let us denote orders of polynomials in $y$ as $N_{P_y}$, $N_{Q_y}$, $N_{X_y}$, $N_{Y_y}$. We see that set of orders of polynomials in equation (\ref{XYyy}) is as follows
\begin{equation}
\{N_{P_y}+2N_{Q_y}+2N_{Y_y}-2;\,3N_{Q_y}+N_{X_y}+N_{Y_y};\,N_{P_y}+2N_{Q_y}+N_{X_y}+N_{Y_y}-1\}.
\label{OrXYyy}
\end{equation}

Examination of equation (\ref{XYyy}) shows that for substantial solutions orders at least of two polynomials there must be equal. It leads to systems of following type
\begin{align}
&N_{P_y}+2N_{Q_y}+2N_{Y_y}-2=N_{P_y}+2N_{Q_y}+N_{X_y}+N_{Y_y}-1;\notag \\ &N_{P_y}+2N_{Q_y}+N_{X_y}+N_{Y_y}-1>3N_{Q_y}+N_{X_y}+N_{Y_y}.
\label{Ord}
\end{align}
The solution of this system is $N_{P_y}-N_{Q_y}>1;\,\, N_{Y_y}-N_{X_y}=1$.

The number of equations in system $N_{Sys}$ here is $N_{P_y}+2N_{Q_y}+2N_{Y_y}-1=N_{P_y}+2N_{Q_y}+N_{X_y}+N_{Y_y}$ and number of independent unknown coefficients $N_{Unc}$ (taking into account possible cancellation of rational function on one nonzero coefficient) is $N_{X_y}+N_{Y_y}+1$. So we conclude that $N_{Sys}-N_{Unc}=N_{P_y}+2N_{Q_y}-1$ and generally speaking (except for $N_{P_y}={0,1}$, $N_{Q_y}=0$ ) $N_{Sys}-N_{Unc}>0$, i.e., there is shortage of unknown coefficients. So a solution of the system exists only if \emph{some conditions (integrability conditions or constrains) for known coefficients of $P$ and $Q$ are hold true}.

Extremely appreciably here that the estimation of \emph{maximal} number of constrains $max N_{Cons}=N_{Sys}-N_{Unc}$ \emph{is not depend on} $N_{X_y}$ and $N_{Y_y}$ at all. Though contents of the constrains strongly depend on these parameters.

We conclude that as in original version of the Prelle-Singer method we are not able here on the base of orders of polynomials define their common degree bound. For many ODEs the degree bound is completely defined by interference properties of \emph{all} coefficients of $P$ and $Q$.

But it is obvious \emph{that the shortage of unknown coefficients here signals first of all that rational generalized integrating factors do not complete the whole family of rational first-order ODEs}.

Considering all possible systems for orders similar to (\ref{Ord}) we can arrange the table of cases for rational $\mu_{yy}$ (Table 1.).

\begin{center}
\begin{tabular}{lccc}
\hline
\hline Case & $N_{P_y}-N_{Q_y}>1$& $N_{P_y}-N_{Q_y}<1$& $N_{P_y}-N_{Q_y}=1$ \\
\hline \\
Conditions on & $N_{Y_y}-N_{X_y}=1$& $N_{Y_y}-N_{X_y}=$& $N_{Y_y}-N_{X_y}\leq1$\\
 $N_{Y_y}$ and $N_{X_y}$ &&$=N_{Q_y}-N_{P_y}+2$&\\
 \\
\hline
\\
$N_{Sys}$ & $N_{P_y}+2N_{Q_y}+$& $3N_{Q_y}+$ &  $N_{P_y}+2N_{Q_y}+$ \\
&$+N_{X_y}+N_{Y_y}$&$+N_{X_y}+N_{Y_y}+1$&$+N_{X_y}+N_{Y_y}$\\
\\
\hline
$max N_{Cons}$& $N_{P_y}+2N_{Q_y}-1$& $3N_{Q_y}$& $3N_{P_y}-3=3N_{Q_y}$\\
\hline
\hline
\\
\multicolumn{4}{c}{Table 1. Order guide for rational $\mu_{yy}$.}
\end{tabular}
\end{center}
\bigskip

Although mathematically all rational generalized integrating factors are equivalent, this does not mean that the actual theirs computations are equally easy.

Similarly we can arrange the tables of cases for rational $\mu_{yx}$, $\mu_{xy}$ and $\mu_{xx}$, (Tables 2-4.).

\begin{center}
\begin{tabular}{lccc}
\hline
\hline Case & $N_{P_y}-N_{Q_y}>1$& $N_{P_y}-N_{Q_y}<1$& $N_{P_y}-N_{Q_y}=1$ \\
\hline \\
Conditions on & $N_{Y_y}-N_{X_y}=$& $N_{Y_y}-N_{X_y}=1$& $N_{Y_y}-N_{X_y}\leq1$\\
 $N_{Y_y}$ and $N_{X_y}$ &$=N_{P_y}-N_{Q_y}$&&\\
 \\
\hline
\\
$N_{Sys}$ & $3N_{P_y}+N_{Q_y}+$& $2N_{P_y}+2N_{Q_y}+$ &  $2N_{P_y}+2N_{Q_y}+$ \\
&$+N_{X_y}+N_{Y_y}$&$+N_{X_y}+N_{Y_y}+1$&$+N_{X_y}+N_{Y_y}+1$\\
\\
\hline
$max N_{Cons}$& $3N_{P_y}+N_{Q_y}-1$&$2N_{P_y}+2N_{Q_y}$& $4N_{P_y}-2=4N_{Q_y}+2$\\
\hline
\hline
\\
\multicolumn{4}{c}{Table 2. Order guide for rational $\mu_{yx}$.}
\end{tabular}
\end{center}
\bigskip

\begin{center}
\begin{tabular}{lccc}
\hline
\hline Case & $N_{P_y}-N_{Q_y}>1$& $N_{P_y}-N_{Q_y}<1$& $N_{P_y}-N_{Q_y}=1$ \\
\hline \\
Conditions on & $N_{Y_y}-N_{X_y}=0$& $N_{Y_y}-N_{X_y}=$& $N_{Y_y}-N_{X_y}\leq0$\\
 $N_{Y_y}$ and $N_{X_y}$ &&$=N_{Q_y}-N_{P_y}+1$&\\
 \\
\hline
\\
$N_{Sys}$ & $2N_{P_y}+2N_{Q_y}+$& $N_{P_y}+3N_{Q_y}+$ &  $2N_{P_y}+2N_{Q_y}+$ \\
&$+N_{X_y}+N_{Y_y}$&$+N_{X_y}+N_{Y_y}+1$&$+N_{X_y}+N_{Y_y}$\\
\\
\hline
$max N_{Cons}$& $2N_{P_y}+2N_{Q_y}-1$&$N_{P_y}+3N_{Q_y}$& $4N_{P_y}-3=4N_{Q_y}+1$\\
\hline
\hline
\\
\multicolumn{4}{c}{Table 3. Order guide for rational $\mu_{xy}$.}
\end{tabular}
\end{center}
\bigskip

\begin{center}
\begin{tabular}{lccc}
\hline
\hline Case & $N_{P_y}-N_{Q_y}>1$& $N_{P_y}-N_{Q_y}<0$& $N_{P_y}-N_{Q_y}=1$ \\
\hline \\
Conditions on & $N_{Y_y}-N_{X_y}=$& $N_{Y_y}-N_{X_y}=0$& $N_{Y_y}-N_{X_y}\leq0$\\
 $N_{Y_y}$ and $N_{X_y}$ &$=N_{P_y}-N_{Q_y}-1$&&\\
 \\
\hline
\\
$N_{Sys}$ & $3N_{P_y}+$& $2N_{P_y}+N_{Q_y}+$ &  $3N_{P_y}+$ \\
&$+N_{X_y}+N_{Y_y}$&$+N_{X_y}+N_{Y_y}+1$&$+N_{X_y}+N_{Y_y}$\\
\\
\hline
$max N_{Cons}$& $3N_{P_y}-1$&$2N_{P_y}+N_{Q_y}$& $3N_{P_y}-1=3N_{Q_y}+2$\\
\hline
\hline
\\
\multicolumn{4}{c}{Table 4. Order guide for rational $\mu_{xx}$.}
\end{tabular}
\end{center}
\bigskip

For some cases there are possibilities to guess a part of structure of rational generalized integrating factors. As a result it reduces an order of polynomial in (\ref{Polyn}) (or, rather, the orders $N_{X_y}$ and $N_{Y_y}$) and somewhat simplifies the problem. Let us consider as an example $\mu_{yy}(x,y)=X(x,y)/(Q(x,y)Y(x,y))$, where $Q$ is the denominator of $f$. Here we get an analog of equation (\ref{XYyy}) in the following form 
\begin{align}
Y^2(PQQ_{yy}-2PQ_y^2&-P_{yy}Q^2+2P_yQQ_y)-Q^2(XY_x+X_x)+\notag \\&+PQ(XY_y-X_y)+X(2PQ_y+QQ_x-P_yQ)=0
\label{QXYyy}
\end{align}
with the following table (Table 5)

\begin{center}
\begin{tabular}{lccc}
\hline
\hline Case & $N_{P_y}-N_{Q_y}>1$& $N_{P_y}-N_{Q_y}<1$& $N_{P_y}-N_{Q_y}=1$ \\
\hline \\
Conditions on & $N_{Y_y}-N_{X_y}=$& $N_{Y_y}-N_{X_y}=$& $N_{Y_y}-N_{X_y}\leq 2-N_{P_y}$\\
 $N_{Y_y}$ and $N_{X_y}$ &$1-N_{Q_y}$&$=2-N_{P_y}$&\\
 \\
\hline
\\
$N_{Sys}$ & $N_{P_y}+N_{Q_y}+$& $2N_{Q_y}+$ &  $N_{P_y}+N_{Q_y}+$ \\
&$+N_{X_y}+N_{Y_y}$&$+N_{X_y}+N_{Y_y}+1$&$+N_{X_y}+N_{Y_y}$\\
\\
\hline
$max N_{Cons}$& $N_{P_y}+N_{Q_y}-1$& $2N_{Q_y}$& $2N_{P_y}-2=2N_{Q_y}$\\
\hline
\hline
\\
\multicolumn{4}{c}{Table 5. Order guide for rational $\mu_{yy}=X/(QY)$.}
\end{tabular}
\end{center}
\bigskip

We can see from Table 5 that here the estimation of maximal number of constrains $max N_{Cons}=N_{Sys}-N_{Unc}$ results in less value.

\section{Simplification of the system for unknown coefficients}

The most natural way for finding unknown coefficients of the structures of rational generalized integrating factors is solving of the system for unknown coefficients, which follows from polynomial equations of type (\ref{Polyn}). For example, for $\mu_{yy}$ this system is determined by the equation (\ref{XYyy}).

Of course, as in every variant of Prelle-Singer procedure from the beginning we have to \emph{assign} a bound to the $N_{X_y}$ (or $N_{Y_y}$). The order of $N_{Y_y}$ (correspondingly $N_{X_y}$) is defined by "Conditions on $N_{X_y}$ and $N_{Y_y}$" from the Tables 1-5.

As we stated above such systems are polynomial ODE-algebraic systems. On the first step here we have to \emph{simplify} this system by any of existent differential elimination procedures \cite {Reid}, \cite {Hubert}\footnote{I have to note here that for the problems described in this paper the \emph{Rif} procedures in \emph{Maple} lead to goal faster.}. The systems described here are not easy by no means. And it is crucial moment of the method. Even for $N_{Sys}=5...6$ the procedure sometimes may hang-up for days.

If we consider ODE with symbolic parameters then under properly (by "Conditions on $N_{X_y}$ and $N_{Y_y}$") assigned $N_{X_y}$ and $N_{Y_y}$ and when $N_{Sys}$ is sufficiently small we always receive a simplified system with some (or without) conditions (constrains) for "known" coefficients of $P$ and $Q$. Then here the ODE solution can be found if these constrains are satisfied and we \emph{are able to solve} simplified system.

In case when ODE is specified without symbolic parameters then we can receive or "System is inconsistent" (and we may augment the bound to the orders of unknown polynomials) or simplified system without any constrains (here we are disquieted if the solution of simplified system can be expressed in some way).

Despite the foregoing and the fact that it is as before a semi-decision procedure, the method described here can solve plenty of unsolved equations.

Compared with pattern matching techniques the method proposed here will certaily resource consuming and so not be so efficient in some cases, but it provides a systematic way of finding solutions without a prior knowledge of what type the equation is. And what is more, in many cases we can use (to pick up speed of the process) pattern matching techniques too with precomputed (by described above prosedures) lists of integrability conditions.

We are not going to give examples here. They take up a lot of space but one easily can generate them, e.g., by means of short \emph{Maple} program like this example for simplest nontrivial ODE: $dy/dx=1/(Y1(x)y+Y0(x))$

\emph{restart;}

\emph{with(PDEtools):}

$N := 0:$

$n[1] := N:$

$n[2] := N+3:$

$f := 1/(Y1(x)*y+Y0(x)):$

$mu := sum(a1[i](x)*y^i,i = 0 .. n[1])/sum(a2[i](x)*y^i,i = 0 .. n[2]):$

$sys := \{coeffs(collect(numer(factor(diff(f,`\$`(y,2))+diff(mu,x)+$

    $+diff(mu,y)*f+diff(f,y)*mu)),y),y,'t')\}:$

$funcs := \{seq(a1[i](x),i = 0 .. n[1]), seq(a2[i](x),i = 0 .. n[2])\}:$

$casesplit(sys,funcs,rif);$

\section{The solution of equation assembling}

If we have found a solution of the system for unknown coefficients (that is we have found, for example, the $\mu_{yy}$) then it remains only to assemble the solution of equation (\ref{ode}). 

We have from (\ref{muyy}) that
\begin{equation}
\zeta(x,y)=F_1(x)+F_2(x)\int e^{\,\int\mu_{yy}(x,y)\,dy}\,dy,
\label{zeta}
\end{equation}
where $F_1(x)$ and $F_2(x)$ have to be determined.
If we substitute (\ref{zeta}) to (\ref{pde}), we can find that
\begin{equation}
F_2(x)=C_2\,\exp\{-\int \mu_{yy} \, dy\}\exp\{-\int(\frac {\partial f}{\partial y}+f\mu_{yy}) dx\}
\label{F2}
\end{equation}
and
\begin{align}
F_1(x)=C_1&-C_2\,\int \exp\{-\int \mu_{yy} \,dy\}\exp\{-\int(\frac {\partial f}{ \partial y}+f\mu_{yy})\, dx\}\times \notag \\&\times\{\int\int\frac {\partial \mu_{yy}}{\partial x}\,dy \,\exp\{\int \mu_{yy}\, dy\}\,dy -\int\exp\{\int \mu_{yy}\, dy\}\,dy\times\notag \\&\times\{\int\frac {\partial \mu_{yy}}{\partial x} \,dy+\frac {\partial f}{\partial y}+f\mu_{yy}\}+f\exp\{\int \mu_{yy}\, dy\}\}\,dx,
\label{F1}
\end{align}
where $C_1$ and $C_2$ are arbitrary constants. Here we may impose that $C_1=0$ and $C_2=1$.

It is easy to verify that $F_1(x)$ and $F_2(x)$ are really functions only in $x$ if $\mu_{yy}$ is a solution of (\ref{muyyeq}).

So, all components of $\zeta$ in (\ref{zeta}) are known now and we can obtain the general solution of equation (\ref{ode}) in implicit form as follows
\begin{equation}
\zeta(x,y)=C,
\label{C}
\end{equation}
where $C$ is an arbitrary constant.

\section{More general structures}

We have concluded above that rational generalized integrating factors do not fit all rational first-order ODE.

There are some possible ways for extension of forms of generalized integrating factors. The most direct of them based on the following idea.

The solution of linear first-order non-homogeneous PDE, e.g.,
\begin{equation}
\frac{\partial \eta}{\partial x}+ f\frac{\partial \eta}{\partial y}+g=0
\label{nheq}
\end{equation}
(where $f=f(x,y)$, $g=g(x,y)$) can be expressed through particular solution of homogeneous PDE
\begin{equation}
\frac{\partial \eta_0}{\partial x}+ f\frac{\partial \eta_0}{\partial y}=0
\label{heq}
\end{equation}
by the following expression
\begin{equation}
\eta(x,y)=F[\eta_0(x,y)]-\int_a^x g(\xi,RootOf[\eta_0(\xi,Z)-\eta_0(x,y)])d\xi,
\label{nheqsol}
\end{equation}
where $F[s]$ is an arbitrary function, $RootOf[\phi(Z)]$ is a root of algebraic equation $\phi(Z)=0$ with respect to $Z$.

Further we can suppose (on the same reasons as explained above) that $\frac{\partial}{\partial y}\ln \frac{\partial \eta_0}{\partial y}$ is rational, then for $g=\frac{\partial f}{\partial y}$ (and $F[s]=s$) from (\ref{muy}) and (\ref{nheqsol}) we obtain a structure for $\mu_{y}=exp(\eta)$, which obviously is more general.

Unfortunately this way leads to sufficiently complicated structures to be analyzed yet.

\section{Rational ODEs only in $x$}

It is obvious that we can consider by the same way first-order ODEs with $f$ rational only in $x$. The corresponding order guides are mirrors of Tables 1-4 under replacement of $P \rightleftarrows Q$ and $x \leftrightarrows y$.

\section{Rational ODEs in both $x$ and $y$}

It is a classical case of the Prelle-Singer method. We can combine different order cases and arrange the order guides too.

Here we seek solution for generalized integrating factors in form
\begin{equation}
\mu_{yy}=\frac{\sum_{i=0}^{N_{X_x}}\sum_{j=0}^{N_{X_y}} a_{ij}x^iy^j}{\sum_{i=0}^{N_{Y_x}}\sum_{j=0}^{N_{Y_y}} b_{ij}x^iy^j},
\label{Polynxy}
\end{equation}
where $a_{ij}$ and $b_{ij}$ are \emph{constants}.

Therefore we obtain the "auxiliary" system in form of \emph{algebraic polynomial system}. The maximal number of constraints here depends on $N_{X_x}$, $N_{X_y}$, $N_{Y_x}$, and $N_{Y_y}$ but we are not able define a common degree bound for polynomial orders too.

\section{Conclusions}

We presented the semi-decision method for solving rational first-order ODEs based on hypothesis that generalized integrating factor of the given ODE is rational function.

Besides first-order ODE the method is suited after some modification for solving some non-linear equations of higher order too.

\end{document}